# Preventive Care Resource Allocation in Developing Countries: Can Rational Planning Techniques Help in Allocating Vaccinators in Dera Ismail Khan District of Pakistan?

Mohammad Ishfaq[1], Faran Ahmad Qadri[1], Nadeem Javaid[2]

[1]King Abdulaziz University, Rabigh, Saudi Arabia.
[2]COMSATS Institute of Information Technology, Islamabad, Pakistan.

**ABSTRACT**
**Preventive care service delivery faces immense challenges when it comes to the level of coverage. Within this context, the case of child immunization is presented by applying operations management tools. This paper explores the application of integer programming techniques to support the Expanded Programme of Immunization (EPI) service in the Dera Ismail Khan District of Pakistan. The main concern here is equitable service delivery to decentralized localities based on two criteria: (1) achieving the highest possible level of vaccination among the target population; and 2) ensuring equality among geographically scattered populations, especially in rural dwellings. For this purpose two integer programming models have been applied on the basis of (1) sub-dividing health district into localities and allocating vaccinators to visit and vaccinate children within their administrative boundaries, and (2) within the localized planning system allowing vaccinators to visit and vaccinate children across the administrative boundaries subject to savings in travel time. Both models show interesting results in terms of need satisfaction and travel-time savings with a minimum level of deviation from equity. The solutions provide a trade-off between alternative organizational tactics, and the argument is made that rational planning methods applied interactively can contribute to the delivery of an immunization service that is equitable and cost effective.**

## 1. INTRODUCTION

There is no doubt that national health is predominantly dependent upon preventive care. This implies that within the domain of primary and secondary health care, more emphasis should be placed on the delivery of preventive care in order to lessen the burden on secondary health care. Health care need is multidimensional, therefore social planners have the difficult task of distributing resources within the social sector that help integrate social sector services, and within health services of subdividing resources appropriately between primary and secondary care. One solution that we have observed in the literature is the decentralization of primary health care services so that the various services complement each other and the benefits reach the grassroots effectively. In one study in Manitoba, Canada, the components of preventive care delivery were examined with reference to three different methods: (1) childhood immunizations (by physicians and public health nurses under a government programme), (2) mammography screening (under a government programme introduced in 1995), and (3) cervical cancer screening (no programme) [1]. The purpose of the Manitoba study was to understand the effect of socioeconomic status on the use of the various components of preventive care, which is different than what we are focusing in this research. The uniqueness of our approach will become clearer as, later in this paper, we will highlight evidence from the literature where the cost of immunization has been the fundamental issue, not the delivery tactics. The study of the impact of socioeconomic status on the use of the service, or the cost behaviour of vaccination, does have the underlying objective of identifying needy groups and delivering immunization services equitably, but tactical planning for service delivery, which is a core issue in this paper, has not been a major concern in the literature.

The main objective of this research is to demonstrate the application of operations management techniques for preventive care planning in the real life scenario of a developing country. In this respect, our paper focuses on the delivery of immunization services through decentralized primary health care in Dera Ismail Khan (DIK) – a health district of Pakistan. However, the scope for decentralized primary care is not the main issue of this paper, as it is evident that decentralization with reference to Pakistan is not a new concept [2]. Like primary care in Manitoba, primary care in the DIK District also runs several programmes in parallel, for example the Lady Health Workers Programme to provide support for child and mother, National Programme for Family Planning, Expanded Programme for Immunization (EPI), Malaria Control Programme, National AIDS Prevention and Control Programme, and Tuberculosis Control Programme. Each of these programmes is delivered through a mixture of door to door visits and clinical surgeries.

As indicated earlier, the major concern in the literature about providing a vaccination service has been the cost of providing the service. For example, a study concentrating on the variations in the cost of delivering a routine immunization service in Peru recommended that its findings about cost variation could be generalized and applied in other settings to support decision makers in their attempts to plan [3]. The main focus of this study was providing disaggregated data on local costs, which planners normally do not have [3]. Our stance, however, is the tactical planning for the equitable delivery of the service that we believe would address differences in cost distribution among localities.

In line with this, the current paper demonstrates the application of advanced planning methods to allocate vaccinators for the Expanded Programme of Immunization (EPI). We have chosen this service firstly because the vaccination service in DIK is provided through door to door visits and it would be possible to apply easily travelling salesman models to this situation, and secondly because preventing infants from catching infectious diseases should be prioritized to give them a good quality life. The use of integer programming models has been demonstrated for the allocation of district nurses to localities in the UK; [4, 5] therefore, in the current case our focus is on showing how integer programming can be used for an analogous situation in a developing country.

According to general perception infectious diseases are a major problem in the developing countries. Thousands of children die of measles, pertussis, poliomyelitis, tuberculosis, diphtheria, and tetanus. Many more are crippled, blinded and spend the rest of their lives with one or more complications of these maladies. Therefore, vaccination is a priority over all other preventive measures to give children a chance of good quality living. So, to vaccinate children under five in Pakistan against the above six diseases, the district health authority deploys vaccinators to visit children in their houses to immunize them under the EPI scheme.

## METHOD

The method we are applying is integer programming, which has been applied elsewhere for similar purposes of providing primary care. Integer programming techniques are widely used for making large-scale strategic planning decisions where circumstances require planning models containing integer-valued variables. Many decision problems involving combinatorial optimization can be solved using integer modelling, such as the deployment of travelling salesmen, machine scheduling, sequencing and so on within resource constraints [6, 7, 8]. Analogous to the travelling salesman approach, the present paper offers two mixed integer programming models to allocate vaccinators for primary health care service delivery in the DIK District.

In order to apply an integer programming model, the input data required are (1) the need for vaccination, and (2) estimated travel time between various need and supply points. The data collection is further explained as follows.

**Need for Vaccination Service**

The method of calculating need for the vaccination service is much simpler than that used for any other primary health care service. It is fairly straight forward because each child that is born has to be vaccinated. Therefore, the 'criterion of need' calculation is based on the population of children. However, it is necessary to establish which age categories are to be considered and how immunization is to be carried out during childhood.

The most common schedule of immunization deals with two age categories of children: the first is 0–1 years and second 4–5 years of age. A standard practice immunization between 0–1 years of age involves five stages of vaccination, which means that each child in this age category has to be visited five times in a year. This implies that the need for immunization for this age category is five times the population of children in this category. However, the need for the immunization of children of 4–5 years of age is simply the same as the total population in that age category.

Based on these criteria, the age-specific need for immunization was worked out on the basis of child population of the union councils of the three localities of DIK [2] and used as input for our models. Administratively, DIK has three localities for health planning: these are Paroa, Dera, and Paharpur, which are further subdivided into union councils, the smallest unit of a district. The need data is not given here to save space.

**Travel Time Estimation**

The second data set required for allocating vaccinators is the estimated travel times between vaccination centres and union councils. In DIK District there are reportedly sixteen vaccination centres which deliver the immunization service throughout 25 union councils (including municipal committee areas as well). The estimated travel times between vaccination centres and union councils were calculated using the DIK road network map. All union councils and the municipal committee areas are interconnected by roads, most of which are metalled. However, some of the union councils are not directly linked by metalled roads, and in such cases unmetalled roads or ordinary tracks are used. In order to estimate travel times we measured distances on metalled roads and on unmetalled and ordinary tracks where appropriate.

Within each union council, most of the population resides within the vicinity of the union council's headquarters, although there are number of small scattered villages nearby. We have therefore estimated the travelling time between union council headquarters and vaccination centres. The method used for estimating travel time is as follows:

1) All the union councils and vaccination centres were located on the road network map.

2) With the help of a map measure, distances between union councils and vaccination centres were measured. For each vaccination centre and union council, distances on metalled and, where appropriate, on unmetalled roads and ordinary tracks were assessed separately in kilometres.

3) Based on these distances motorcycle travelling times were calculated. Motorcycle travelling time was taken because most vaccinators are provided with motorcycles for this purpose (although within the city vaccinators may walk or use bicycles).

4) Assuming an average speed of 30 km/hour on metalled roads and 10 km/hour on unmetalled or ordinary tracks, travel times on the two road surface types were obtained separately in minutes.

5) Finally the travel times on metalled roads and unmetalled or ordinary tracks were combined to give total travel time in minutes between each vaccination centre and each union council. A matrix of travel times between vaccination centres and union councils was developed and used as input data for both the models. This matrix is not given here to save space.

**Assumptions**

The data on need and travel times were then used to carry out an exercise to allocate vaccinators to localities using the following assumptions:

1) The decentralized service may be delivered in two ways, namely (i) as a strictly locality-bound service, and (ii) as a flexible service allowing service across the board. This has allowed us to use two models. In Model 1 we assumed that there are three localities in DIK, each comprising a set of vaccination centres and union councils. It was assumed that the service delivery would take place strictly within the prescribed locality boundaries – cross-boundary flows were disallowed. Model 2 considers that localities serve primarily as managerial units for the set of vaccination centres located within them; however, the vaccination centres may serve any set of union councils throughout the tehsil (called locality in our case. For administrative purposes district in Pakistan is sub divided into tehsils, which in turn are further sub divided into union councils) – allowing cross boundary flows. The objective of analysing the situation in two different ways was to examine the implications of removing locality-based restrictions for equitable service delivery.

2) On an average a vaccinator can conveniently visit and vaccinate five children per day. The number of children assumed to be vaccinated per day was lower than expected because vaccinators have to visit children in their homes and because of travelling difficulties and poor roads within each union council. Using this assumption we can see that both models suggest that, with the current level of provision, the need that can be satisfied will remain below 59%. Looking at the current health situation in Pakistan this result seems quite reasonable and would provide a basis for EPI to cover more of the population.

3) There are 273 working days in a year (365 less 52 weekly days off and 40 public holidays and other leave).

4) Travel time between vaccination centres and union councils has been taken as a measure of cost.

The explanation of the two models is given below.

**Model One**

This model focuses on service deliver strictly within the context of existing localities, and it aims to allocate an integer number of vaccinators to each locality so that the proportion of need met is as nearly equal as possible across all union councils and across all children to be vaccinated (the equity constraint), and so that the total travelling time of vaccinators is minimized. A full description of the model is given in Appendix 1.

**Model Two**

This model assumes that localities remain the managerial units for the vaccination centres within their boundaries, but vaccinators from any vaccination centre can be allocated to serve the populations of any union council within the district. This assumption has been made to examine the implications of removing locality-based restrictions on service delivery. The aim is the same as in Model 1, to allocate an integer number of vaccinators to each locality in such a way that the proportion of the need that is met is the same (the equity constraint), and that the total time travelled by vaccinators is minimized. A full description of Model 2 is given in Appendix 2.

# RESULTS

**Model One**

In the case of Model 1, where the service is delivered strictly within the context of proposed locality boundaries, a feasible solution was provided when a 3% deviation from equity was allowed. The model was then tested repeatedly allowing deviations from equity of 5%, 10%, 15%, 20%, and 25% in order to examine the impact on the allocation of vaccinators to localities and on travel time. A summary of results is given in Table 1.

Table 1: Results from the solution of Model One with various equity levels to allocate vaccinators in DIK

| Annual Travel Time (in hours) | 22370 | 220988 | 21376 | 20529 | 19781 | 19132 |
|---|---|---|---|---|---|---|
| Deviation from Equity | 3% | 5% | 10% | 15% | 20% | 25% |
| $\alpha_{max}$ | 58.6% | 59.5% | 61.9% | 62.6% | 63.9% | 65.6% |
| $\alpha_{min}$ | 55.6% | 54.5% | 51.9% | 47.6% | 43.9% | 40.6% |
| Number of vaccinators in each locality | | | | | | |
| 1) Paroa | 16 | 16 | 15 | 15 | 15 | 15 |
| 2) Dera | 12 | 12 | 13 | 13 | 13 | 13 |
| 3) Paharpur | 18 | 18 | 18 | 18 | 18 | 18 |

$\alpha_{max}$ and $\alpha_{min}$ are measure of maximum and minimum need that is satisfied.

Table 1 reveals that, with a 3% deviation from equity, the total immunization need that could be satisfied for each union council and each age category was within the range of 55.6% to 58.6%. The table also shows that as the deviation from equity increases, the distance travelled reduces; so permitting equity to vary by 25% rather than 3% results in a 14.5% saving in travel time. Otherwise, it is evident that the overall allocation pattern to localities does not show any major change.

Although 14.5% is a considerable reduction in travel time, it must be set against the importance of child immunization. The two dimensions are incommensurable, but priority must clearly be given to the latter, suggesting that vaccinators should be allocated with a minimum feasible deviation from equity.

This model has the capability to allocate vaccinators to the localities and individual vaccination centres to serve each union council and each age category, but for the sake of simplicity we have not gone into such details here.

**Model Two**

This model considers the allocation of vaccinators in a situation where localities comprise only a set of vaccination centres which can serve any set of union councils throughout the tehsil. The results obtained from the solution for Model Two are given in Table 2.

Table 2: Results from the solution of Model Two to allocate vaccinators in DIK Pakistan

| | |
|---|---|
| Annual travelling time in hours: | 19,192 |
| $\alpha$ (a measure of equitable need satisfied): | 57.1% |
| Number of vaccinators by localities | |
| 1) Paroa | 11 |
| 2) Dera | 16 |
| 3) Paharpur | 19 |

Table 2 shows that the travelling time in Model 2 is 14.2% less than that for Model 1 with a 3% equity deviation. The level of need satisfied equitably is 57.1% in each union council and in each age category. This means that by disregarding locality boundaries and allowing cross boundary flows there may be a reasonable saving in travel time.

**Discussions**

Both models are similar in that their objective is to allocate vaccinators to localities. However, they solve the problem by assuming two different methods of service delivery. Model 1 disallows cross boundary flows whereas Model 2 relaxes this condition. The models were run using the mathematical programming system Sciconic and the solutions obtained are explained below.

Both Model 1 and Model 2 have the capability to allocate vaccinators to localities and vaccination centres serving each age category in each union council. For the sake of simplicity this data is not presented here; however, it is worth mentioning that given the proposed localities, this solution allows some cross boundary flows. In order to reach a decision about whether to implement the Model 1 or 2 solution, interaction with the responsible decision-making body would be required. It is interesting to observe that the solution obtained by Model 2 in particular is quite close to actual pattern of vaccinator deployment on the basis of proposed localities, as can be seen in Table 3.

The integer programming models used in this study have investigated useful alternatives for the rational allocation of vaccinators for preventive care in DIK. From the above results we are of the view that rational planning methods applied interactively can contribute to the delivery of an immunization service that is equitable and cost effective. Such an approach can be easily generalized and applied in any situation where service delivery requires visiting the recipients of the service, and it also allows planners to find the trade-offs between alternative organizational tactics, as we have shown in case of decentralized care.

Our discussion would be incomplete if we do not raise the issue of data reliability in the case of developing countries, and Pakistan is not the only case where data collection is a major impediment to the application of rational planning tools. Nevertheless, developing countries are making progress in this area as, for example, we see efforts being made to establish a Health Management Information System in Pakistan [9, 10]. From the literature it is also evident that there are adequate sources of data with reference to health, for example, the Pakistan Medical Research Council (www.pmrc.org.pk), the Aga Khan University (www.aku.edu) and many others in public sector, which would enable analysts to develop models with a reasonable confidence.

Our work with regard to preventive care planning is only indicative; a fuller study could be conducted interactively with the responsible decision-making body, perhaps with access to sources of more accurate data. We have also shown here how an operations managers can have flexibility in their decision making and find trade-offs between equality of service delivery and cost savings, provided the options are established in consultation with the local community as well as in coordination with the central authority.

## CONCLUSIONS

This paper reports on the work – the first of its kind – to demonstrate the impact of applying a strategic planning technique to the effective and equitable delivery of preventive care (child immunization) in the DIK district of Pakistan. The allocation pattern resulting from the integer programming models was not out of line with the actual deployment of vaccinators at the time of data collection.

With the increasing emphasis on preventive care in Pakistan and elsewhere [1, 11] there is increasing evidence of costs being taken into consideration, whereas our work shows cost effectiveness can be achieved by shifting towards localized planning for the delivery of services. Localized planning will evolve over a period of time, although there may be a number of issues to be tackled. Our work here is just a preliminary step in the introduction of rational planning methods which are used in developed countries but which it is equally possible to apply in developing countries.

The scope of this work does not end here. Looking at the sub-continent region we can see the latest statistics on health presented in India's Annual Report [12]. There is awareness in the subcontinent of the importance of the use of information in decision making. Expertise is available in the region and countries can benefit from mutual cooperation. The subcontinent region, which includes India, Bangladesh, Sri Lanka and Nepal, all have the similar population characteristics with a variety of talents, technologies and resources available. These can be applied effectively in rational planning and strategic decision making in the social sector to help develop the respective nations by enhancing the quality of life through the provision of adequate and effective preventive care. There is also no shortage of intellectual and material input from the West. Researchers [13] and organizations like the UN and South Pacific Partnership are playing their part in uplifting the preventive care system of developing countries by diagnosing problems, sharing information and providing resources. It is necessary to take advantage of this by deploying resources in the right direction.

This research opens the door to the development of operations management techniques with reference to social sector planning in the area of health in developing countries, and it adds knowledge about operations management techniques that can be used as useful tools for planning social sector services in developing countries.

ACKNOWLEDGEMENT

The authors are grateful to Professor Jonathan Rosenhead of LSE, Susan Powell of LSE and Professor Ann Taket of Deakin University for their invaluable help and guidance in developing the models. The authors are also grateful to Dr Abdulelah Saaty, Dean of the College of Business, Rabigh, King Abdulaziz University, KSA for his help and support in conducting this research.